\documentclass[twocolumn,nopacs,amsmath,amssymb,superscriptaddress,prl]{revtex4}
\usepackage{graphicx}
\usepackage{dcolumn}
\usepackage{bm}
\usepackage{color}
\usepackage{braket}

\newcommand{\1}{\uparrow}
\newcommand{\2}{\downarrow}
\newcommand{\be}{\begin{equation}}
\newcommand{\ee}{\end{equation}}

\begin{document}
\title{Beyond-mean-field effects in Rabi-coupled two-component Bose-Einstein condensate}

\author{L. Lavoine}
\affiliation{Laboratoire Charles Fabry, UMR 8501, Institut d'Optique, CNRS, Universit\'e Paris-Saclay, Avenue Augustin Fresnel, 91127 Palaiseau CEDEX, France}
\author{A. Hammond}
\affiliation{Laboratoire Charles Fabry, UMR 8501, Institut d'Optique, CNRS, Universit\'e Paris-Saclay, Avenue Augustin Fresnel, 91127 Palaiseau CEDEX, France}
\author{A. Recati}
\affiliation{INO-CNR BEC Center and Dipartimento di Fisica, Universit\`a degli Studi di Trento, 38123 Povo, Italy and Trento Institute for Fundamental Physics and Applications, INFN, 38123 Trento, Italy}
\author{D.S. Petrov}
\affiliation{Universit\'e Paris-Saclay, CNRS, LPTMS, 91405 Orsay, France}
\author{T. Bourdel}
\email[Corresponding author: ]{thomas.bourdel@institutoptique.fr}
\affiliation{Laboratoire Charles Fabry, UMR 8501, Institut d'Optique, CNRS, Universit\'e Paris-Saclay, Avenue Augustin Fresnel, 91127 Palaiseau CEDEX, France}

\date{\today}
\begin{abstract}
We theoretically calculate and experimentally measure the beyond-mean-field (BMF) equation of state in a coherently-coupled two-component Bose-Einstein condensate (BEC) in the regime where averaging of the interspecies and intraspecies coupling constants over the hyperfine composition of the single-particle dressed state predicts the exact cancellation of the two-body interaction. We show that with increasing the Rabi-coupling frequency $\Omega$, the BMF energy density crosses over from the nonanalytic Lee-Huang-Yang (LHY) scaling $\propto n^{5/2}$ to an expansion in integer powers of density, where, in addition to a two-body BMF term $\propto n^2 \sqrt{\Omega}$, there emerges a repulsive three-body contribution $\propto n^3/\sqrt{\Omega}$. We experimentally evidence this two contributions, thanks to their different scaling with $\Omega$, in the expansion of a Rabi-coupled two-component $^{39}$K condensate in a waveguide. By studying the expansion with and without Rabi coupling, we reveal an important feature relevant for observing BMF effects and associated phenomena in mixtures with spin-asymmetric losses: Rabi coupling helps preserve the spin composition and thus prevents the system from drifting away from the point of vanishing mean field.
\end{abstract}

\maketitle

Vacuum effects are among the most striking features of quantum field theories. The high degree of control of cold gases has made these systems ideal candidates to identify and measure the role of quantum fluctuations in matter fields. For instance, using Bose gases, quantum phonons fluctuations \cite{ArmijoFluc}, phononic Lamb shift \cite{OberthalerLamb}, dynamical phononic Casimir effect \cite{WestbrookCasimir}, as well as evidence of Hawking-like radiation from analog gravity configurations \cite{Steinhauer} have been  reported. 

For a weakly interacting Bose gas quantum fluctuations cause the BEC depletion \cite{Chang2016, Lopez2017} and lead to the so-called LHY correction to the mean-field (MF) equation of state \cite{LHY}. This BMF correction has found experimental verification a few years ago \cite{CornellLHY,SalomonLHY}.  More recently, self-trapped quantum droplets were stabilized against collapse by BMF effects in single component dipolar BECs \cite{PfauDroplet,FerlainoDroplet} and in two-component BEC mixtures \cite{Petrov2015, Cabrera2018, Semeghini2018, Guo2021}. In the latter case, the MF interaction is greatly reduced by the compensation between intraspecies repulsion and interspecies attraction while the BMF energy originating from quantum fluctuations remains finite. 

In this Letter we consider a Bose gas formed by atoms with two internal levels, which are coherently coupled \cite{Goldstein1997}. The BEC order parameter is then a two-component vector and the relative phase excitations are gapped due to the coherent drive. Such systems have been throughly analyzed on the MF level and used in variety of experiments on coherent Rabi oscillations \cite{RabiCornell99,RabiCornell2000}, internal self-trapping effects \cite{Zibold2010}, ferromagnetic classical bifurcation  \cite{Nicklas2011}, Kibble-Zurek mechanism \cite{OberthalerKZ}, control of the two-body interaction \cite{Hanna2010, Petrov2014, Sanz2019}, and magnetic domain wall dynamics \cite{Farolfi2020}. For sufficiently attractive interspecies interaction such Rabi-coupled BECs are predicted to sustain droplets in the symmetric case, where the Rabi coupling is resonant and the intraspecies coupling constants are equal to each other. The stabilization mechanism is interpreted from the few-body perspective as an emergent three-body repulsion \cite{Petrov2014} or as a many-body BMF effect due to a structural change in the excitation spectrum \cite{Cappellaro2017}. 

Here, we analytically calculate the BMF energy density in the experimentally relevant case of an asymmetric Rabi-coupled Bose mixture in the whole range of Rabi frequencies $\Omega$. We show that the BMF energy crosses over from the LHY law $\propto n^{5/2}$ for small $\Omega$ to the regular expansion in integer powers of $n$ for large $\Omega$. The quadratic term in the latter limit can be understood as a renormalization of the two-body interaction, the cubic term as an emergent three-body interaction, each term having a different scaling with $\Omega$. In the large $\Omega$ limit, we experimentally evidence these two contributions by preparing a condensed $^{39}$K spin-mixture at the point of vanishing mean field \cite{Skov2020} and by quantitatively measuring its expansion in a waveguide as a function of $\Omega$. 
In the absence of Rabi coupling, we observe that asymmetric losses between the two spin states drive the spin composition away from the point of zero MF interaction and thus strongly affect the expansion.

We consider a Bose gas of $N$ atoms possessing two internal states, $\sigma=\1,\,\2$, coherently coupled with the Rabi frequency $\Omega$ and detuning $\delta$. In the rotating wave approximation the Hamiltonian of the mixture reads \cite{Goldstein1997,RecatiAbadEPJD}
\begin{eqnarray}\label{Ham}
\hat{H}&=& \int\left(\sum_{\sigma} -\hat{\Psi}^\dagger_{\sigma {\boldsymbol{r}}}\frac{\nabla^2_{\boldsymbol{r}}}{2}\hat{\Psi}_{\sigma {\boldsymbol{r}}}\right. +\sum_{\sigma\sigma'}\hat{\Psi}^\dagger_{\sigma {\boldsymbol{r}}} \xi_{\sigma\sigma'} \hat{\Psi}_{\sigma' {\boldsymbol{r}}}
\nonumber\\ 
&+&\left.\sum_{\sigma,\sigma'}
 \frac{g_{\sigma\sigma'}}{2}  \hat{\Psi}^\dagger_{\sigma {\boldsymbol{r}}} 
 \hat{\Psi}^\dagger_{\sigma' {\boldsymbol{r}}}\hat{\Psi}_{\sigma {\boldsymbol{r}}} 
 \hat{\Psi}_{\sigma' {\boldsymbol{r}}}\!\!\right)d{\bf r},
\end{eqnarray} 
where $\hat{\Psi}_{\sigma {\boldsymbol{r}}}$ is the annihilation Bose field operator, $\xi=-(\delta\sigma_z+\Omega\sigma_x)/2$ is the single particle spin Hamiltonian written in terms of the Pauli matrices acting in the $\ket{\1}$-$\ket{\2}$ space, and $g_{\sigma\sigma'}=4\pi a_{\sigma\sigma'}$ are the coupling constants for the $\sigma$-$\sigma'$ interaction with the scattering lengths $a_{\sigma\sigma'}$. In Eq.~(\ref{Ham}) and in the rest of the paper we adopt the units $\hbar=m=1$, where $m$ is the mass of the particles.

Assuming zero-temperature and weak interactions, we follow the usual Bogoliubov procedure by separating the dominant condensate contribution and writing the field operator as $\hat{\Psi}_{\sigma {\boldsymbol{r}}}=\sqrt{n_\sigma}+\hat{\phi}_{\sigma {\boldsymbol{r}}}$, where $n_\sigma$ are the condensate densities and $\hat{\phi}_{\sigma {\boldsymbol{r}}}$ annihilate particles with nonzero momenta. Neglecting $\hat{\phi}$ and substituting $\hat{\Psi}_{\sigma {\boldsymbol{r}}}=\sqrt{n_\sigma}$ into Eq.~(\ref{Ham}) we obtain the MF energy density
\begin{equation}
E_{\rm MF}=\frac{\delta(1-\alpha^2)-2\Omega\alpha}{2(1+\alpha^2)} n+\frac{g_{\1\1}\alpha^4+g_{\2\2}+2g_{\1\2}\alpha^2}{(1+\alpha^2)^2}\frac{n^2}{2},\label{EnergyVSalpha}
\end{equation}
where $n=n_\1+n_\2$ and $\alpha=\sqrt{n_\1/n_\2}$. In the limiting case of vanishing density (or interactions), $E_{\rm MF}$ gets minimized for $\alpha=\alpha_0=\delta/\Omega+\sqrt{1+\delta^2/\Omega^2}$, consistent with the condensation in the dressed state $\ket{-}=(\alpha_0 \ket{\1} +\ket{\2})/\sqrt{1+\alpha_0^2}$, which is the ground state of $\xi$. The coefficient in front of $n^2/2$ in Eq.~(\ref{EnergyVSalpha}) is then the MF coupling constant corresponding to the scattering length $a_{--}=(a_{\1\1}\alpha_0^4+a_{\2\2}+2a_{\1\2}\alpha_0^2)/(1+\alpha_0^2)^2$ \cite{Search01, Sanz2019}. For $a_{\1\1}>0$, $a_{\2\2}>0$, and $a_{\1\2}<0$, $a_{--}$ exhibits a minimum as a function of $\delta$ (or $\alpha_0$). We are interested in the particular configuration of the scattering lengths (controlled by the magnetic field) and the RF drive parameters where this minimum touches zero. This translates into two conditions: 
\begin{equation}\label{ThePoint}
\delta a=0 \textrm{ and } \alpha_0=\beta,
\end{equation} 
where we have introduced the scattering length detuning $\delta a=a_{\1\2}+\sqrt{a_{\1\1}a_{\2\2}}$ and the interaction asymmetry parameter $\beta=(a_{\2\2}/a_{\1\1})^{1/4}$. Note that for finite $n$ the energy $E_{\rm MF}$, upon minimization with respect to $\alpha$, is not, in general, a quadratic function of $n$ since the optimal polarization parameter $\alpha$ does depend on $n$. In particular, the system can feature a three-body attraction already on the MF level~\cite{Hammond21}. However, the minimum of $a_{--}$ is a special point where both terms on the right-hand side of Eq.~(\ref{EnergyVSalpha}) are minimized at $\alpha=\alpha_0$ independent of $n$. At this point $E_{\rm MF}$ is thus quadratic in $n$. If, in addition, we tune $\delta a$ to zero [configuration (\ref{ThePoint})], the minimum of $a_{--}$ also vanishes and the condensate becomes noninteracting on the MF level.

We shall now discuss the influence of BMF effects on the equation of state. In the Bogoliubov approach the leading BMF term is obtained by expanding the Hamiltonian (\ref{Ham}) up to quadratic terms in $\hat{\phi}_\sigma$ and by summing the zero point energies of the corresponding Bogoliubov modes. In the symmetric case ($a_{\1\1}=a_{\2\2}$ and $\delta=0$) the calculation has been performed in Ref.~\cite{Cappellaro2017}. The general asymmetric case is technically more difficult because of cumbersome expressions for the Bogoliubov modes \cite{SM}. However, under the conditions (\ref{ThePoint}), these expressions simplify and read
\begin{align}
&E_{p,-}=p^2/2,\nonumber \\
&E_{p,+}=\sqrt{(p^2/2+\tilde{\Omega})
(p^2/2+\tilde{\Omega}-2g_{\1\2}n)},\nonumber
\end{align}
where $p$ is the momentum and $\tilde{\Omega}=\Omega(\alpha_0+1/\alpha_0)/2$. The BMF energy density can then be reduced to the form~\cite{SM}
\begin{equation}\label{LHY}
E_{\rm BMF}=\frac{8(-g_{\1\2}n)^{5/2}}{15\pi^2} I\left(\frac{\tilde{\Omega}}{-2g_{\1\2}n}\right),
\end{equation}
where $I(y)=(15/4)\int_0^1\sqrt{x(1-x)(x+y)}dx$. Equation~(\ref{LHY}) remains a good approximation for the BMF energy density as long as $|\delta a/a_{\1\2}|\ll 1$ and $|\alpha_0-\beta|\ll 1$. The function $I(y)$ is a monotonically growing function, which tends to 1 for $y=\tilde{\Omega}/{(-2g_{\1\2}n)}  \rightarrow 0$. This is the limit of two uncoupled condensates where Eq.~(\ref{LHY}) reduces to the LHY form, responsible for the BMF stabilization of quantum droplets in binary mixtures \cite{Petrov2015}. In the opposite limit $I(y)$ can be expanded in powers of $1/y\ll 1$, the first two leading terms being $I(y)\approx (15\pi/128)(4\sqrt{y}+1/\sqrt{y})$. The substitution of this expansion into Eq.~(\ref{LHY}) gives
\begin{align}\label{LHY3}
E_{\rm BMF}&\approx \frac{\sqrt{\tilde{\Omega}}}{2\sqrt{2}\pi}
g_{\1\2}^2 \frac{n^2}{2}+{\frac{3}{{4\sqrt{2}\pi\sqrt{\tilde{\Omega}}}}}
|g_{\1\2}|^{3}\frac{n^3}{6},
\end{align}
which is qualitatively different from the LHY $n^{5/2}$ scaling. The two terms in Eq.~(\ref{LHY3}) can be interpreted as a BMF-renormalized two-body interaction and an emergent three-body term, both with a specific scaling with $\Omega$. Interestingly, both these terms agree with exact two-body \cite{SM} and three-body calculations \cite{3b} in the limit $\sqrt{\tilde{\Omega}}|g_{\1\2}|\ll 1$. Higher-order terms require taking into account interactions between excited (noncondensed) atoms and are thus not captured by the quadratic Bogoliubov Hamiltonian. For the same reason, Eq.~(\ref{LHY3}) contains no LHY term $\propto n^{5/2}$, expected to be ``induced'' by the BMF-renormalized two-body interaction. This term, however, is of very high order $\propto (\sqrt{\tilde{\Omega}}g_{\1\2}^2n)^{5/2}$ and is negligible for our experimental parameters.

In our experiment we directly measure the BMF energy at the point of  vanishing MF through the released energy in a one-dimensional expansion. We work with the second and third lowest Zeeman states of the lowest manifold of $^{39}\rm{K}$, namely  $\ket{\1}=\ket{\rm{F}=1,\rm{m_{F}}=-1}$ and $\ket{\2}=\ket{\rm{F}=1,\rm{m_{F}}=0}$. At a magnetic field of $56.830(1)\,\rm{G}$, the three relevant scattering lengths are $a_{\1\1}=33.4\,a_0$, $a_{\2\2}=83.4\,a_0$ and  $a_{\1\2}=-53.2\,a_0$, where $a_0$ is the atomic Bohr radius \cite{Tiemann2020}. The minimum of $a_{--}$ is then $-0.2\,a_0$. We have checked that the corresponding residual MF energy is a small correction as compared to the BMF energy for our parameters. The experiment starts with a quasi pure BEC in state $\ket{\1}$. The atoms are optically trapped in an elongated harmonic potential  with frequencies $(\omega_x,\omega_y,\omega_z)=(137, 137,25.4)\; \rm{Hz}$.

The coherent mixture in $\ket{-}$ is prepared in an adiabatic passage, in which the radio-frequency (RF) detuning is swept from $\delta=7.5\, \Omega$ to its final value $\delta \approx 0.23\,\Omega$, for which $\alpha_0\approx \beta$. During the RF sweep, $a_{--}$ decreases from $33.3\,a_0$ to a value close to $0$. Its shape and duration of 9\,ms are chosen in order to be adiabatic not only with respect to the internal-state dynamics but also with respect to the radial evolution of the wave function, which progressively shrinks and approaches the ground state of the radial harmonic confinement. We have checked that we cannot detect any residual oscillations of the radial size after the RF sweep. In addition, the axial trap frequency is progressively turned off during the sweep and adjusted such that the gas is neither axially expanding nor shrinking in a three-dimensional expansion after the sweep. The radially confined condensate is then free to expand in the axial direction for a time of 75\,ms. During the final stage of the expansion, we sweep back the RF frequency to its initial value to map the dressed states $\ket{-}$ and $\ket{+}$ back onto $\ket{\1}$ and $\ket{\2}$, which we independently detect by fluorescent imaging after a short Stern-Gerlach separation. 

Interestingly, we observe that the atoms remain in state $\ket{-}$ \cite{losses}, which might be surprising as losses are known to take place mostly in state $\ket{\2}$ \cite{Semeghini2018, Cheiney2018} and thus could lead to the creation of $\ket{+}$. In order to understand the spin dynamics, we can model the losses by adding $-i\Gamma/2$ to the second diagonal term in $\xi$. Diagonalizing the resulting matrix, we see that for $\Gamma\ll \Omega$ the initial state $\ket{-}$ remains essentially unchanged, except that it decays with the rate $\approx \Gamma |\braket{-|\2}|^2=\Gamma/(1+\alpha_0^2)$. In our experiment $\Gamma$ is at most $\sim20\,$s$^{-1}$, which is much lower than $\Omega$. We thus conclude that the spin-dependent loss in our case reduces to an effective loss in the dressed state, with $\Gamma$ weighted by the fraction of the lossy component.

\begin{figure}[!th]
\centering
       \includegraphics[width=0.95\columnwidth, clip]{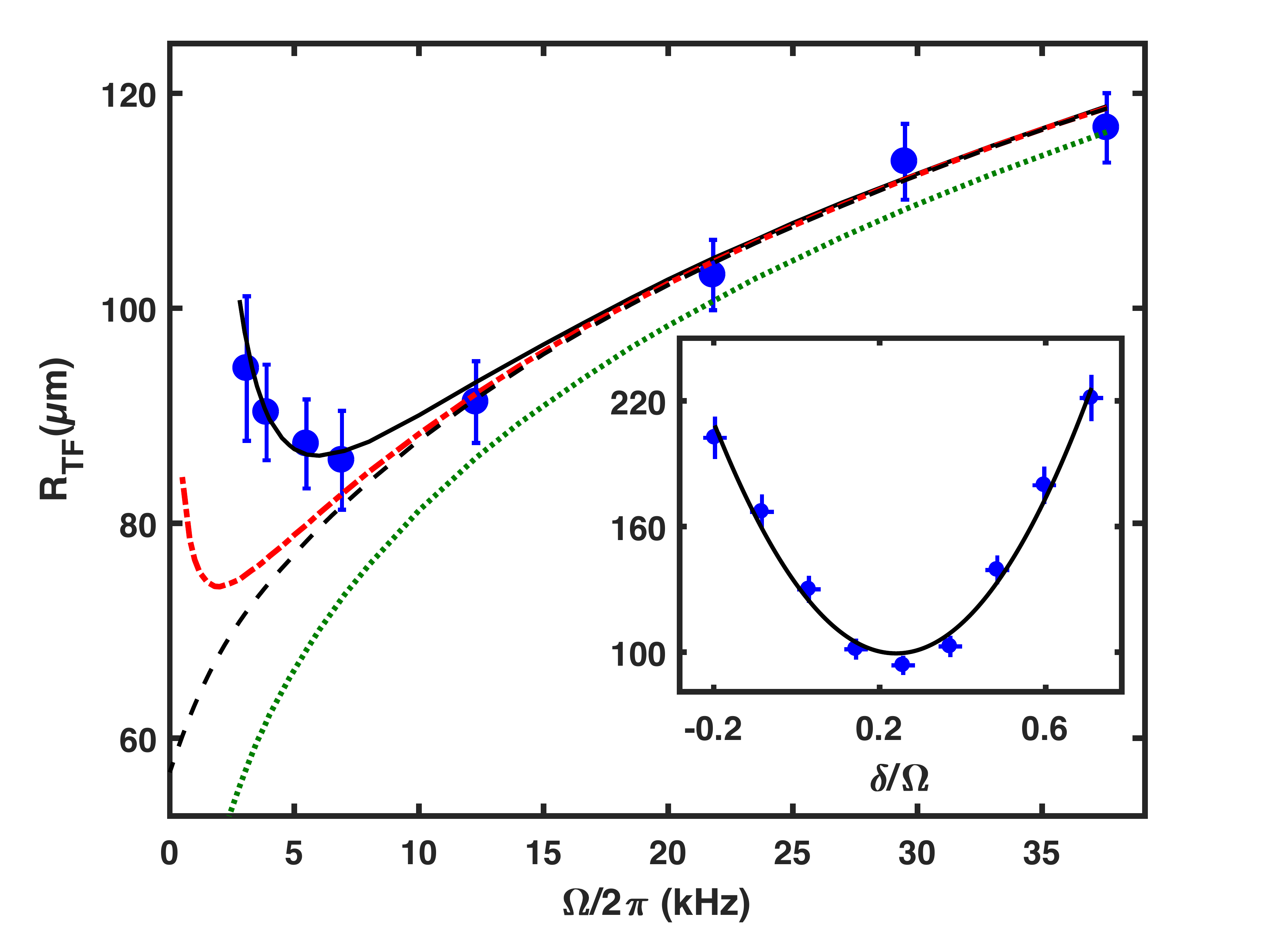}
        \caption{Minimal Thomas-Fermi radius of the condensate after $75\,\rm{ms}$ of expansion as a function of the Rabi coupling strength (for $\delta/\Omega=0.23$). Each point is obtained by averaging 15 fluorescence images and the error bars correspond to the single shot standard deviation. The curves correspond to quasi-1D extended Gross-Pitaevskii simulations (see text) with or without magnetic field noise and assuming different forms of the BMF term. Black solid curve: full BMF term with noise; black dashed curve: full BMF without noise; green dotted curve: two-body BMF term without noise; red dash-dotted curve: two-body plus three-body BMF term without noise. Inset: Experimental Thomas-Fermi radius of the condensate after $75\,\rm{ms}$ of expansion as a function of $\delta/\Omega$, for a Rabi frequency $\Omega/2\pi=12.29\,\rm{kHz}$. The curve is a parabolic guide to the eye.\label{LHYexp}}
\end{figure} 

For each value of the Rabi frequency $\Omega$, we first measure the condensate density profile after expansion and extract the condensate sizes from fits with 1D Thomas-Fermi density profiles $\propto 1-(z/R_\textrm{TF})^2$ (see inset of Fig.\,\ref{LHYexp}). As a function of the detuning $\delta$, a clear minimum in size appears at $\delta/\Omega \approx 0.23$, i.e., a position where we expect $a_{--}$, and thus also the MF energy to be minimized \cite{BMFvar}. As explained above, under the conditions (\ref{ThePoint}), the MF term vanishes at its minimum ($a_{--}\approx 0$) and the expansion of the cloud is governed by the BMF term. In Fig.\,\ref{LHYexp}, we plot the measured minimal size as a function of the Rabi frequency $\Omega/2\pi$. For Rabi frequencies $\Omega/2\pi$ between 6\,kHz and 38\,kHz, we observe a slow increase of the measured size as a function of $\Omega$, which corresponds to an increase of the BMF energy as predicted previously. More precisely, the size after a long time of flight is expected to scale with $\sqrt{E^{\rm}_{\rm LHY}}\propto \Omega^{1/4}$ for a dominant two-body term at large $\Omega$, in qualitative agreement with our observed behavior. At low Rabi frequencies, below 6\,kHz, we observe an increase of the size which we attribute to low frequency magnetic field noise. In a 50\,$\mu$s Ramsey sequence, we have measured a standard deviation of the magnetic field $\Delta B\sim0.8(2)\,$mG corresponding to $\Delta\delta \sim 560(140)\,$Hz. These fluctuations lead to an increased average value of the scattering length $\Delta a_{--}\approx 48 (\Delta \delta/\Omega)^2 \,a_0$, which can explain our observed increasing sizes for low values of $\Omega$ \cite{average}. 

In order to precisely model our experiment we solve the wave-function evolution through a single component one-dimensional non-linear Schr\"odinger equation in which we properly account for the BMF energy $E_{\rm BMF}$ \cite{GPELab}. The axial initial wave function is taken to be $\propto 1-(z/R_\textrm{TF})^2$, such that the axial density profile is the one of the initial 3D Thomas-Fermi condensate before the RF sweep. The spin healing length is smaller than the radial cloud size and the BMF energy is treated in a local density approximation in the radial direction \cite{Lavoine2021}. Since the chemical potential can be of order the radial confinement energy $\omega_\perp$ (especially at large $\Omega$), we take into account, as a function of the 1D density, the correction to the Gaussian radial profile due to the dominant two-body BMF contribution. The 1D energy density is then found by radial integration, and the chemical potential entering the Sch\"odinger equation by partial derivation with respect to the 1D density. The magnetic field noise is accounted for by an increase of $a_{--}$ entering the MF term.  

As can be seen from Fig.~\ref{LHYexp} this model quantitatively reproduces our experimental data. The black curves are generated by using Eq.~(\ref{LHY}) and assuming the measured finite $\Delta B$ (solid) and $\Delta B=0$ (dashed). For this calculation the initial atom number $1.05\times 10^5$ has been adjusted to match the experimental data and corresponds within 5\% to a calibration using the condensation temperature (determined with $\sim$20\% accuracy). The initial peak density is $n\sim 5\times 10^{14}\,$cm$^{-3}$ and the three-body loss coefficient $K_{---}/3!=4\times10^{-28}\,$cm$^6$s$^{-1}$ has been adjusted to match our observed $\sim 30\%$ atom loss during the expansions \cite{K3}.

The other curves in Fig.~\ref{LHYexp} are obtained by running the same code with $\Delta B=0$ using Eq.~(\ref{LHY3}) (red dash-dotted) or restricting it only to the two-body term (green dotted). They show that in our explored range $\Omega/2\pi>3\,$kHz the BMF energy is very well approximated, at least theoretically, by the sum of the dominant two-body contribution $E_{\rm BMF}\propto \sqrt{\Omega}$ and a smaller three-body term $\propto 1/\sqrt{\Omega}$. To give concrete numbers, for $\Omega/2\pi=10$\,kHz, $\tilde{\Omega}/(-2g_{\1\2}n)\approx 1$ and the initial two-body and three-body energies per particle are calculated to be 20\,Hz and  3.2\,Hz, respectively. 

In order to better understand how much of this theory is experimentally tested, we first note that the scaling $E_{\rm BMF}\propto \sqrt{\Omega}$ is clearly evidenced in our data for $\Omega/2\pi>6\,$kHz. Assuming that the next-order term scales as $\propto 1/\sqrt{\Omega}$ we try to verify if the corresponding coefficients are consistent with Eq.~(\ref{LHY3}). To this end we fit the data with the prediction of our dynamical model based on Eq.~(\ref{LHY3}) with two free parameters: the total atom number and a prefactor $\eta$, inserted by hand in Eq.~(\ref{LHY3}) in front of the three-body term. Repeating this fitting procedure for different values of $\Delta B$ we find $\eta=0.85(35)$, where the error bar is dominated by the uncertainty in our determination of the magnetic field noise. Our measurements as a function of $\Omega$ are thus in quantitative agreement with Eq.~(\ref{LHY3}). The two terms,  named two-body and three-body according to their predicted density scalings, are independently measured from their specific $\Omega$ dependence.

\begin{figure}[!th]
\centering
\includegraphics[width=0.9\columnwidth, clip=true]{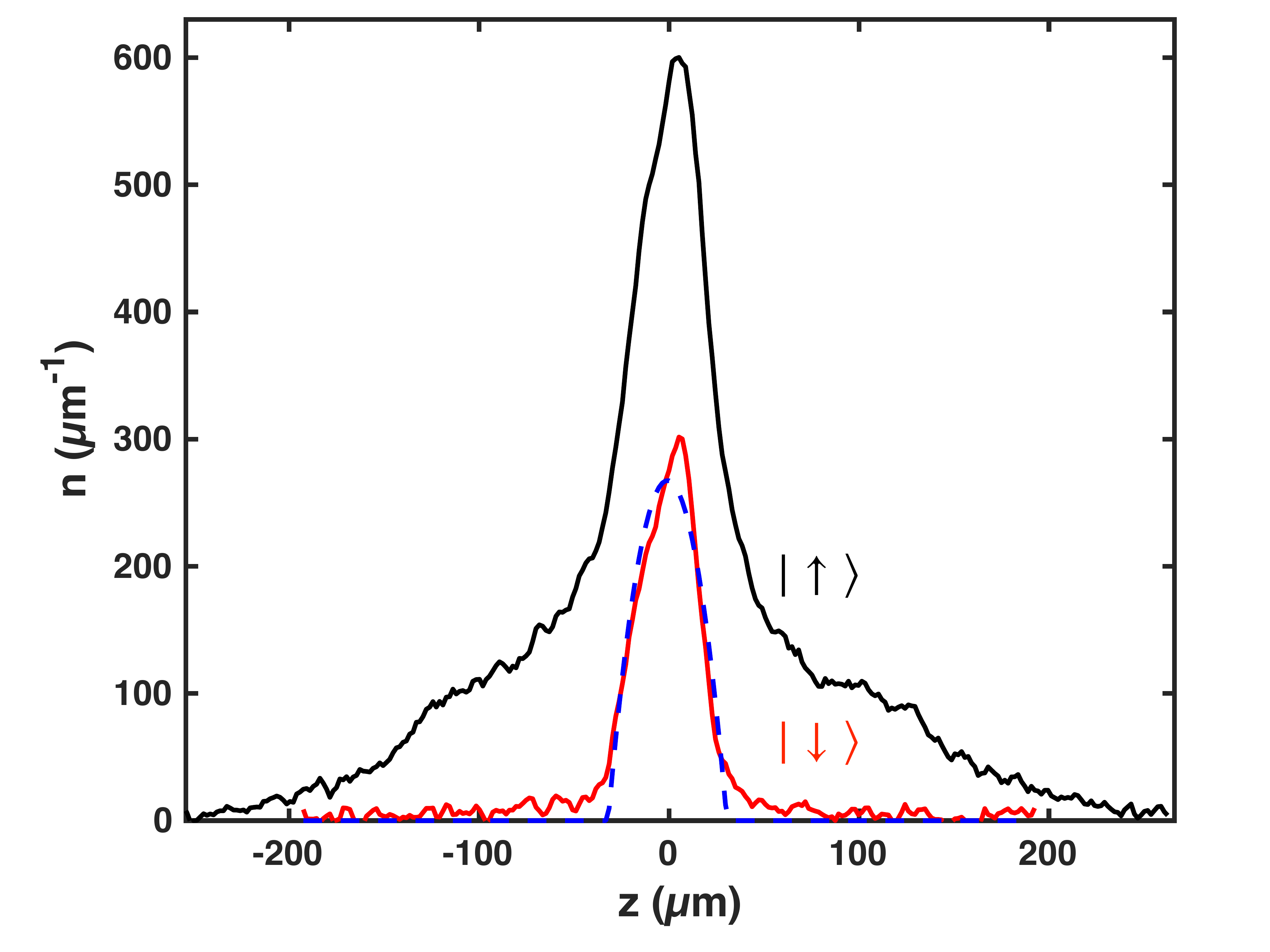}
\caption{Density profiles for $\ket{\1}$ (black) and $\ket{\2}$ (red) atoms after 75\,ms of expansion at $\Omega=0$. The blue dashed curve is a 1D Thomas-Fermi fit of the density profile in state $\ket{\2}$.}
\label{incoherent}
\end{figure} 

Finally, we can also repeat the expansion measurement while removing the RF coupling field at the end of the sweep such that we are left with two uncoupled condensates. The losses then dominantly take place in state $\ket{\2}$ and contrary to the coupled case, $\alpha$ quickly deviates from its initial value $\beta$. Moreover, we observe after a Stern-Gerlach separation that the two clouds behave differently in the expansion (see Fig.\,\ref{incoherent}). The $\ket{\2}$ condensate does not expand much whereas the $\ket{\1}$ condensate exhibits a double structure with a low energy central part. This behavior is reminiscent of previous observations in droplet configuration where excess $\ket{\1}$ atoms are expelled from the droplet region \cite{Cabrera2018, Semeghini2018, Ferioli2020}. In addition, we find that the condensate 1D Thomas-Fermi radius in state $\ket{\2}$ is 31\,$\mu$m, a value that is significantly lower than the expected radius of 57\,$\mu$m for our parameters in the single component simulation with $\Omega \rightarrow 0$, i.e. with the BMF energy density scaling with $n^{5/2}$. This difference indicates a significant role of transient MF effects in the expansion dynamics of the central region. Here, $\ket{\1}$ atoms, which are more abundant than expected and which require some time to escape, create an excessive effective trapping for $\ket{\2}$ atoms forcing their slower expansion.

In conclusion, we have studied both theoretically and experimentally, the BMF equation of state of a coherently-coupled two-component BEC in the asymmetric case. The BMF correction is most manifest when the spin polarization and interactions are tuned to the special point given by Eqs.~(\ref{ThePoint}) where MF effects are minimized. The BMF energy density as a function of $n$ then interpolates between the usual $\propto n^{5/2}$ LHY form in the uncoupled limit to a qualitatively different behavior in the strong-coupling regime where one can introduce effective BMF two-body and three-body interactions with specific scalings as a function of the Rabi coupling strength $\Omega$. The experimentally measured condensate expansions are governed by the BMF energy and are in good agreement with the theory at large $\Omega$. We detect not only the dominant two-body  BMF term $\propto \sqrt{\Omega}$, but also evidence the smaller three-body  term $\propto 1/\sqrt{\Omega}$, opening the path to the creation of coherently-coupled quantum droplets in which the two-body interaction (MF+BMF) is compensated by BMF three-body effects \cite{Bulgac2002, Petrov2014}. 
Exploration of the small $\Omega$ regime will require a further reduction of the magnetic field fluctuations. Interestingly, a coherent Rabi coupling helps to preserve the spin composition and thus prevents the system from dynamically drifting away from the point of vanishing mean field and thus facilitates direct measurements of the BMF equation of state. 

\begin{acknowledgments}
We thank L. Tarruell for useful discussions. This  research  has  been  supported  by  CNRS,  Minist\`ere  de  l'Enseignement  Sup\'erieur  et  de  la  Recherche, Labex PALM, Region Ile-de-France  in  the  framework  of  DIM  Sirteq, Paris-Saclay in the framework of IQUPS, ANR Droplets (19-CE30-0003), Simons foundation (award number 563916:  localization of waves). AR acknowledge financial support from Provincia Autonoma di Trento, the FISh project of the Istituto Nazionale
di Fisica Nucleare, and the Italian MIUR under the
PRIN2017 project CEnTraL.
\end{acknowledgments}


\begin{thebibliography}{0}
\bibitem{losses}
We have found that  at low Rabi frequencies ($\Omega/2\pi\leq7\,$kHz), magnetic field noise around these frequencies transfers a small fraction of the atoms to state $\ket{+}$ during the condensate expansion (3\% at $\Omega/2\pi=6.9\,$kHz, 4.5\% at $\Omega/2\pi=5.5\,$kHz, 10\% at $\Omega/2\pi=3.9\,$kHz, 12\% at $\Omega/2\pi=3.1\,$kHz).
\bibitem{BMFvar}
In principle, the condensate expansion is governed by the total energy $E^{\rm}_{\rm MF}+E_{\rm BMF}$. However, the variation of the MF energy as a function of $\delta$ is much stronger than the corresponding variation of $E_{\rm BMF}$ such that the latter can be approximated by a constant. As a consequence, the variation of our measured size is dominated by the variation of $a_{--}$. We have checked that for our parameters including the $\delta$ dependence of $E_{\rm BMF}$ leads to shifts of the minimal energy below $1\%$.
\end{thebibliography}

\end{document}


\title{Supplemental material: Beyond-mean-field effects in Rabi-coupled two-component Bose-Einstein condensate}

\author{L. Lavoine}
\author{A. Hammond}
\author{A. Recati}
\author{D.S. Petrov}
\author{T. Bourdel}



\maketitle

\renewcommand{\theequation}{S\arabic{equation}}
\renewcommand{\thefigure}{S\arabic{figure}}

\setcounter{equation}{0}
\setcounter{figure}{0}

\section{Bogoliubov spectrum and BMF energy in coupled binary mixtures}

The expansion of the Hamiltonian~(1) (hereafter, equation numbers without prefix S refer to the main text) up to quadratic terms in the fields $\hat{\phi}_{\sigma,{\bf p}}$ (we switch to momentum space) can be written as $E_{\rm MF}+E_0+\hat{H}_2$, where $E_{\rm MF}$ is given by Eq.~(2),
\begin{widetext}
\be\label{E0'}
E_0=\frac{1}{2}\sum_{\bf p}\left(-p^2-\frac{n_\1+n_\2}{\sqrt{n_\1 n_\2}}\frac{\Omega}{2}-g_{\1\1}n_\1 - g_{\2\2}n_\2 +\frac{1}{p^2}\sum_{\sigma\sigma'}g^2_{\sigma\sigma'}n_\sigma n_{\sigma'}\right),
\ee
and
\be\label{HamQuadratic}
\hat{H}_2=\!\frac{1}{2}(\hat{\phi}_\1^\dagger \hat{\phi}_\2^\dagger \hat{\phi}_\1 \hat{\phi}_\2)\!
\begin{pmatrix}

\hspace{-0.0cm}\frac{p^2}{2}+g_{\1\1}n_\1+\frac{\Omega}{2}\sqrt{\frac{n_\2}{n_\1}} & g_{\1\2}\sqrt{n_\1 n_\2}-\frac{\Omega}{2} & g_{\1\1}n_\1 & g_{\1\2}\sqrt{n_\1 n_\2} \\

g_{\1\2}\sqrt{n_\1 n_\2}-\frac{\Omega}{2} & \hspace{-0.0cm}\frac{p^2}{2}+g_{\2\2}n_\2+\frac{\Omega}{2}\sqrt{\frac{n_\1}{n_\2}} & g_{\1\2}\sqrt{n_\1 n_\2} & g_{\2\2}n_\2 \\

g_{\1\1}n_\1 & g_{\1\2}\sqrt{n_\1 n_\2} & \hspace{-0.0cm}\frac{p^2}{2}+g_{\1\1}n_\1+\frac{\Omega}{2}\sqrt{\frac{n_\2}{n_\1}} & g_{\1\2}\sqrt{n_\1 n_\2}-\frac{\Omega}{2} \\

g_{\1\2}\sqrt{n_\1 n_\2} & g_{\2\2}n_\2 & g_{\1\2}\sqrt{n_\1 n_\2}-\frac{\Omega}{2} & \hspace{-0.0cm}\frac{p^2}{2}+g_{\2\2}n_\2+\frac{\Omega}{2}\sqrt{\frac{n_\1}{n_\2}} 

\end{pmatrix}
\begin{pmatrix}
\hat{\phi}_\1 \\
\hat{\phi}_\2 \\
\hat{\phi}_\1^\dagger \\
\hat{\phi}_\2^\dagger
\end{pmatrix}.
\ee
The first four terms in the brackets on the right-hand side of Eq.~(\ref{E0'}) arise as a compensation for the ``incorrectly'' ordered terms added to the quadratic form (\ref{HamQuadratic}) in order to make it symmetric with respect the ordering of creation and annihilation operators. This symmetrized form is convenient since it stays symmetrized under the usual Bogoliubov transformation and diagonalizes into
\begin{equation}\label{HamQuadraticDiag}
\hat{H}_2=\frac{1}{2}\sum_{{\bf p},\pm} E_{p,\pm}(\hat{b}_{{\bf p},\pm}^\dagger \hat{b}_{{\bf p},\pm}+ \hat{b}_{{\bf p},\pm} \hat{b}_{{\bf p},\pm}^\dagger)=\sum_{{\bf p},\pm} E_{p,\pm}\hat{b}_{{\bf p},\pm}^\dagger \hat{b}_{{\bf p},\pm} + \frac{1}{2}\sum_{{\bf p},\pm} E_{p,\pm}.
\end{equation}
The last term in Eq.~(\ref{E0'}) is due to the standard renormalisation of the coupling constant $g_{\sigma\sigma'}\rightarrow g_{\sigma\sigma'}(1+g_{\sigma\sigma'}\sum_p 1/p^2)$ in the MF term.

The dispersion relations $E_{p,\pm}$ of the  Bogoliubov modes (created by operators $\hat{b}^\dagger_{{\bf p},\pm}$) can be written as
\begin{equation}\label{BogEn}
E_{p,\pm}=\sqrt{D_p\pm\sqrt{D_p^2-\frac{p^2}{2}\left(\frac{p^2}{2}+\frac{\Omega}{2}\frac{n_\1 +n_\2}{\sqrt{n_\1 n_\2}}\right)\left[\prod_\sigma\left(\frac{p^2}{2}+2g_{\sigma\sigma}n_\sigma+\frac{\Omega}{2}\sqrt{\frac{n_{\bar{\sigma}}}{n_\sigma}}\right)-\left(2g_{\1\2}\sqrt{n_\1 n_\2}-\frac{\Omega}{2}\right)^2\right]}},
\end{equation}
where
\begin{equation}\label{D}
D_p=\frac{1}{2}\sum_\sigma\left(\frac{p^2}{2}+\frac{\Omega}{2}\sqrt{\frac{n_{\bar{\sigma}}}{n_\sigma}}\right)\left(\frac{p^2}{2}+2g_{\sigma\sigma}n_\sigma+\frac{\Omega}{2}\sqrt{\frac{n_{\bar{\sigma}}}{n_\sigma}}\right)-\frac{\Omega}{2}\left(2g_{\1\2}\sqrt{n_\1 n_\2}-\frac{\Omega}{2}\right).
\end{equation}


From Eq.~(\ref{D}) one sees that the mode $E_{p,-}$ is gapless, while $E_{p,+}$ has a gap $\sqrt{2D_0}\neq 0$. The former is just the Goldstone mode due to the breaking of the $U(1)$ symmetry in the condensed state, while the latter is related to the gap introduced by $\Omega$ in having a different phase for the two spinor components. The gap vanishes for $\Omega=0$, since in this case there exist two Goldstone modes of the broken $U(1)\times U(1)$ symmetry. 

The BMF energy is obtained by adding the vacuum part of Eq. (\ref{HamQuadraticDiag}) to the constant energy $E_0$. The BMF energy thus explicitly reads 
\begin{equation}\label{LHY_complete}
E_{\rm BMF}=\frac{1}{2}\int {\frac{d^3p}{(2\pi)^3}} \left(E_{p,+}+E_{p,-}-p^2-\frac{n_\1+n_\2}{\sqrt{n_\1 n_\2}}\frac{\Omega}{2}-g_{\1\1}n_\1 - g_{\2\2}n_\2 +\frac{1}{p^2}\sum_{\sigma\sigma'}g^2_{\sigma\sigma'}n_\sigma n_{\sigma'}\right),
\end{equation}
\end{widetext}
where we have replaced the sum over momenta by the integral. Equation~(4) is obtained from Eq.~(\ref{LHY_complete}) under the conditions (3) as follows. We switch to the integration variable $z=p^2$ and express the integral in Eq.~(\ref{LHY_complete}) as a contour integral around the branch cut of $\sqrt{z}$ along the positive real semiaxis. We then deform this contour towards the negative semiaxis such that it now goes around the branch cut of $E_{p,+}$. This is a finite interval which we map onto $x\in [0,1]$.

\section{Exact solution of the two-body problem in the zero-range approximation}

In order to characterize the two-body scattering let us introduce the single-particle states (normalization will not be important) $\ket{+}=\ket{\1}-\alpha_0\ket{\2}$ and $\ket{-}=\ket{\1}+\ket{\2}/\alpha_0$, which are eigenstates of the operator $\xi$ corresponding to eigenvalues $\pm\sqrt{\Omega^2+\delta^2}/2$. For two bosons we thus have three possible ``total spin'' states, which correspond to three scattering channels equally separated in energy by $\sqrt{\Omega^2+\delta^2}$. We define these spin states by
\begin{eqnarray}
\ket{--}&=&\ket{\1\1}+(\ket{\1\2}+\ket{\2\1})/\alpha_0+\ket{\2\2}/\alpha_0^2,\label{mm}\\
\ket{+-}&=&\ket{\1\1}-(\ket{\1\2}+\ket{\2\1})\delta/\Omega-\ket{\2\2},\label{pm}\\
\ket{++}&=&\ket{\1\1}-\alpha_0(\ket{\1\2}+\ket{\2\1})+\alpha_0^2\ket{\2\2}.\label{pp}
\end{eqnarray}

Consider the scattering of two ground-state $\ket{-}$ bosons at the collision energy $q^2$ below the gap $\sqrt{\Omega^2 + \delta^2}$. Since the atoms interact only at zero separation, the orbital wave functions in the three channels correspond to free motion, respectively, at energies $q^2 > 0$ (open channel), $q^2-\sqrt{\Omega^2 + \delta^2}<0$, and $q^2-2\sqrt{\Omega^2+ \delta^2}<0$ (these channels are therefore closed). In addition, only $s$ waves are important. Accordingly, introducing $\kappa_1=\sqrt{\sqrt{\Omega^2 + \delta^2}-q^2}$ and $\kappa_2=\sqrt{2\sqrt{\Omega^2 + \delta^2}-q^2}$ we write the two-body scattering state as
\begin{equation}\label{ansatz}
\left[\frac{\sin(qr)}{qr}-a\frac{\cos(qr)}{r}\right]\ket{--}+C_1\frac{e^{-\kappa_1 r}}{r}\ket{+-}+C_2\frac{e^{-\kappa_2 r}}{r}\ket{++},
\end{equation}
where the parameters $a$, $C_1$, and $C_2$ are fixed by three zero-range Bethe-Peierls boundary conditions ensuring that for $r\rightarrow 0$ the projections of Eq.~(\ref{ansatz}) on the spin states $\ket{\1\1}$, $\ket{\1\2}+\ket{\2\1}$, and $\ket{\2\2}$ are proportional, respectively, to $1-a_{\1\1}/r$, $1-a_{\1\2}/r$, and $1-a_{\2\2}/r$. Note that $C_1$, $C_2$, and the overall prefactor in Eq.~(\ref{ansatz}) absorb the normalization coefficients omitted in Eqs.~(\ref{mm}-\ref{pp}). Performing the projection leads to the matrix equation
\begin{equation}\label{matrix}
\begin{bmatrix}
1/a_{\2\2} & -\alpha_0^2(\kappa_1-1/a_{\2\2}) & \alpha_0^4 (\kappa_2-1/a_{\2\2})\\
1/a_{\1\2} & -\alpha_0(\delta/\Omega)(\kappa_1-1/a_{\1\2}) & -\alpha_0^2 (\kappa_2-1/a_{\1\2})\\
1/a_{\1\1} & \kappa_1-1/a_{\1\1} & \kappa_2-1/a_{\1\1}\\
\end{bmatrix}
\begin{pmatrix}
a\\C_1\\C2
\end{pmatrix}
=
\begin{pmatrix}
1\\1\\1
\end{pmatrix}.
\end{equation}
The energy-dependent scattering length $a(q)$ is related to the $s$-wave scattering amplitude by the equation
\begin{equation}\label{scatampl}
f(q)=-\frac{1}{1/a(q)+iq}
\end{equation}
and can be obtained from Eq.~(\ref{matrix}) by eliminating $C_1$ and $C_2$. The corresponding explicit expression is analytic but bulky. However, in the limit of small $a_{\sigma\sigma'}$ one can recover the MF result $a(0)=a_{--}$, where $a_{--}$ is defined in the main text. If one assumes $a_{\1\2}\sim a_{\2\2} \sim |a_{\1\2}|$ and $|\delta|\sim \Omega$, one can see from Eq.~(\ref{matrix}) that $a(0)$ can be expanded in powers of the small parameter $|a_{\1\2}|\sqrt{\Omega}\ll 1$. The leading BMF correction to $a$ is thus expected to be of order $a_{\1\2}^2\sqrt{\Omega}$. By solving Eq.~(\ref{matrix}) exactly under the conditions Eq.~(3) we obtain  
\begin{equation}\label{aexact}
a(0)=\sqrt{2}a_{\1\2}^2\sqrt{\tilde{\Omega}}\frac{1+a_{\1\2}\sqrt{\tilde{\Omega}}}{1+[1-2\sqrt{2}+\sqrt{2}(\alpha_0^2+1/\alpha_0^2)]a_{\1\2}\sqrt{\tilde{\Omega}}-\sqrt{2}(\alpha_0-1/\alpha_0)^2a_{\1\2}^2\tilde{\Omega}},
\end{equation}
the leading-order asymptote of which is indeed proportional to $a_{\1\2}^2\sqrt{\Omega}$ (since $a_{--}=0$) and corresponds to the two-body coupling constant in Eq.~(5).  

The term $\propto a_{\1\2}^2\sqrt{\Omega}$ can be understood as the second-order Born correction to the scattering amplitude for small $a_{\sigma\sigma'}\sqrt{\Omega}$. This term accounts for virtual interaction-induced transitions of two atoms in the unperturbed ground $\ket{--}$ state with zero momenta to various spin states ($\ket{--}$, $\ket{+-}$, $\ket{++}$) and momenta ${\bf q}$ and $-{\bf q}$. In spite of the fact that the channels $\ket{+-}$ and $\ket{++}$ are strongly gapped at large $\Omega$ the second-order integral actually grows as $\sqrt{\Omega}$. This can be understood by looking at the structure of this integral, which behaves (qualitatively) as $\int g^2 d^3q/(q^2+\Omega)$. Its diverging part $\int g^2 d^3q/q^2$ is regularized in the standard manner by replacing the bare coupling constants $g_{\sigma\sigma'}$ by $4\pi a_{\sigma\sigma'}$ and the remaining part indeed scales as $\sqrt{\Omega}$ as the number of momentum states essentially contributing to the integral grows faster than the gap.

Finally, we note that the Bogoliubov quadratic Hamiltonian, which leads to the BMF energy Eq.~(4), does take into account the virtual two-body processes which we have just described. Equation~(5) therefore exactly reproduces the second-order Born term. On the other hand, the third-order two-body correction $\propto a_{\1\2}^3\Omega$ [present in Eq.~(\ref{aexact})] is not captured on the level of the Bogoliubov Hamiltonian [and we therefore do not see it in Eq.~(5)] since it requires taking into account interactions between atoms excited from the condensate.
